\documentclass[aps,prd,onecolumn,superscriptaddress]{revtex4}
\usepackage{graphicx}
\usepackage{amsmath}
\usepackage{epsfig, epstopdf}
\usepackage{amsmath,amsfonts,amssymb}
\usepackage[hypertex]{hyperref}

\usepackage{color}

\begin{document}
\baselineskip=12pt
\def\black{\textcolor{black}}
\def\red{\textcolor{black}}
\def\blue{\textcolor{blue}}
\def\green{\textcolor{black}}
\def\be{\begin{equation}}
\def\ee{\end{equation}}
\def\bea{\begin{eqnarray}}
\def\eea{\end{eqnarray}}
\def\orc{\Omega_{r_c}}
\def\om{\Omega_{\text{m}}}
\def\E{{\rm e}}
\def\bearst{\begin{eqnarray*}}
\def\eearst{\end{eqnarray*}}
\def\peleven{\parbox{11cm}}
\def\peffec{\peight{\bearst\eearst}\hfill\peleven}
\def\pspace{\peight{\bearst\eearst}\hfill}
\def\ptwelve{\parbox{12cm}}
\def\peight{\parbox{8mm}}

\title{Simultaneous effect of  Modified Gravity  and  Primordial Non-Gaussianity in Large Scale Structure Observations  }

\author{Nareg Mirzatuny}
\email{mirzatuny-AT-physics.sharif.edu}
\address{Department of Physics, Sharif University of
Technology, P.O.Box 11365--9161, Tehran, Iran}

\author{Shahram Khosravi}
\email{khosravi-AT-mail.ipm.ir}
\address{Department of Physics, Kharazmi University, Tehran, Iran}
\address{School of Astronomy, Institute for Research in
Fundamental Sciences (IPM),
P.~O.~Box 19395-5531,
Tehran, Iran}

\author{Shant Baghram}
\email{baghram-AT-ipm.ir}
\address{School of Astronomy, Institute for Research in
Fundamental Sciences (IPM),
P.~O.~Box 19395-5531,
Tehran, Iran}

\author{Hossein Moshafi}
\email{hosseinmoshafi-AT-iasbs.ac.ir}
\address{Department of Physics, Institute for Advanced Studies in Basic Science (IASBS),
P.O.Box 45195-1159, Zanjan, Iran}
\vskip 1cm

\begin{abstract}
In this work we study the simultaneous effect of primordial non-Gaussianity and the modification of the gravity in $f(R)$ framework on large scale structure observations. We show that  non-Gaussianity and modified gravity introduce a scale dependent bias and growth rate functions. The deviation from $\Lambda$CDM  in the case of primordial non-Gaussian models is in large scales, while the growth rate deviates from $\Lambda$CDM in small scales for modified gravity theories. We show that the redshift space distortion can be used to distinguish  positive and negative $f_{NL}$ in standard background, while in $f(R)$ theories they are not easily distinguishable. The galaxy power spectrum is generally enhanced in presence of non-Gaussianity and modified gravity. We also obtain the scale dependence of this enhancement. Finally we define galaxy growth rate and galaxy growth rate bias as  new observational parameters to constrain cosmology.

\end{abstract}

\maketitle


\section{INTRODUCTION}

The  modern cosmology is now in a dilemma, on one hand the standard model of cosmology known as $\Lambda$CDM describes a vast range of observations, from the Cosmic Microwave Background (CMB) radiation \cite{Komatsu:2010fb} to the observations  related to the  Large Scale Structures (LSS) of the Universe  \cite{Tegmark:2003ud,Tegmark:2003uf,Eisenstein:2005su}. On the other hand $95\%$ of the Universe is made of the unknowns (Dark energy (DE)$\sim 68\%$ and Dark Matter (DM) $\sim 27\%$)\cite{Ade:2013zuv}. Beside this,  the physics of early universe, despite the successes of inflationary paradigm \cite{Guth:1980zm} is also still unknown.

   Regarding the accelerated expansion of the  Universe, there are three main  categories of  probable solutions: a) The Cosmological Constant (CC) \cite{Carroll:2000fy}; b) Dark Energy models \cite{Peebles:2002gy};  and c) The Modified Gravity (MG) theories  \cite{Nojiri:2006ri,Sotiriou:2008rp, DeFelice:2010aj}.

    In the context of the early universe cosmology, the deviation from the  1) scale invariant primordial power spectrum \cite{Adams:2001vc}, 2) the statistical isotropy of the  perturbations \cite{Erickcek:2008sm}, 3) the  adiabatic perturbations  \cite{Gordon:2000hv} and 4) Gaussian initial conditions \cite{Bartolo:2004if} are under an intense study in order to test the inflationary models. Any observed deviation from the above conditions will open a new horizon to study the physics of inflation.

    Now the cosmological observations and the vast data obtained from ground based telescopes and satellites open up a new era to test the cosmological models. One important question is the existence of the simultaneous effects of early universe physics and the late time cosmology on observational parameters. The LSS observations can be both affected from inflationary models and dark energy models.
    In this work we want to mainly address to these simultaneous effects on LSS.

 The clustering and the statistics of LSS of the Universe (i.e. galaxies, cluster of galaxies, voids) have long been known as a useful tool to constrain the cosmology \cite{Seljak:2004sj}. However to constrain cosmological models, there are three type of obstacles  in using the LSS observations:

 ~ a){\it{ Non-linear structure formation}}: In scales corresponding to the wavenumber
 $k\sim 0.2 h/Mpc$ and larger wavenumbers (smaller scales), the structure formation process becomes non-linear \cite{Scoccimarro:2000gm}. Consequently we can not use the linear perturbation theory to study the cosmological models and we need semi-analytical models \cite{Smith:2002dz} to probe further, or use the N-body simulations\cite{Springel:2005nw}.  In this work {\it {we mainly focus on the wavenumber interval of $0.01 h/Mpc < k< 0.2 h/Mpc$ where the perturbations are in the linear/quasi-linear regime}}. In these scales we also have statistically significant data from clustering of galaxies \cite{Ahn:2012fh}.

 b){\it{ Redshift space distortion (RSD)}}: The second complication becomes from RSD effect. Observationally,  instead of measuring the radial coordinate, we measure redshift of the sources through spectroscopy. The redshift measurement in its turn is affected by the peculiar velocity of the structures and it is mixed up with the Hubble expansion redshift and adds more complications to interpret the cosmological results from LSS observations \cite{Hatton:1997xs,Seljak:2000jg}.
 However it can be used as a measure of matter density, where in this work we use RSD as an observational probe for MG theories.

 c) {\it{Bias}}: The third obstacle in LSS observations is the bias parameter. Bias parameter in linear order is defined as the ratio of density contrasts of the luminous matter ($\delta_g$) and the underlying dark matter ($\delta_m$) as ($b=\delta_g/\delta_m$). In observations, what we measure is the clustering of luminous matter which has to be related to the underlying dark matter halo distribution, which in its turn must be related to dark matter density perturbation. This is a complicated process affected by the baryon physics, non-linear structure formation, galaxy formation and evolution. However this parameter is used to study the effect of primordial NG on the distribution of matter in Universe.

  ~~ To be more precise, the above mentioned complications are themselves useful probes for studying the cosmological models. The redshift space distortion is used as a probe to measure the growth rate of the structures, $f\equiv d\ln\delta/d\ln a$, to underpin the expansion history of the Universe and to test the gravitational law in cosmological scales.  It is difficult to measure the growth rate parameter straightforwardly in observations. Instead it is obtained via the knowledge of bias parameter and the redshift space distortion parameter $\beta\equiv f/b$ \cite{Kaiser:1987qv}
obtained in galaxy power spectrum/correlation functions \cite{Blake:2011rj}.
 On the other hand the bias parameter is recently introduced \cite{Dalal:2007cu} as a new probe to detect the fingerprint of the primordial non-Gaussianity in LSS. Primordial NG introduces a scale dependent behavior in the bias parameter.

 {\it{The main question is that how the LSS observables will be affected by MG and NG, when we have these deviations from $\Lambda$CDM both in the same time.}}
 In this work we study the effect of MG and primordial non-Gaussianity(NG) on growth rate and RSD parameters. The main point here is that: It has been shown that the growth of the structures in MG theories (i.e. $f(R)$) is scale  dependent \cite{Bertschinger:2008zb,Tsujikawa:2009ku}. This scale dependence is manifested in the growth rate. Consequently the growth rate can change the redshift space distortion and can be a potentially good observation to test the gravity \cite{Raccanelli:2012gt}.
  We assert that if our universe has a slight deviation from a CC dominated Universe and also a small local NG, $f_{NL}\sim 5$ (which parameterized the strength of the NG) allowed by recent Planck data \cite{Ade:2013ydc}, then the growth rate and the bias parameter will have a non trivial scale dependence. These features  will affect the galaxy power spectrum measurement due to redshift space distortion.  This will cause to a simultaneous effect between cosmological parameters and primordial NG \cite{Carbone:2010sb} worth to study in detail. We chose the specific model of $f(R)$ introduced by Hu-Sawicki (HS)\cite{Hu:2007nk} to show the discussed effects and we show the relation of our chosen model with more general parameterizations of MG introduced by Bertschinger and Zukin (BZ) \cite{Bertschinger:2008zb}.  In this direction, Parfrey, Hui and Sheth propose the emergence of scale dependence in MG will affect the linear bias \cite{Parfrey:2010uy}. We assume that the linear bias modification due to MG is smaller than the MG effect on NG-bias. Also a  very recent study, consider the effect of NG and MG in 3D correlation function of matte power spectrum\cite{Raccanelli:2013dza}.

The structure of this work is as follows: In Sec. (\ref{Sec-Mg}) we introduce the $f(R)$  modified gravity theories and discuss the modified matter growth rate and its scale dependence. In Sec.(\ref{Sec-bias}) we study the bias parameter with primordial NG and the deviation from $\Lambda$CDM. In Sec.(\ref{Sec-galaxy}), we study  the redshift space distortion parameter and the galaxy power spectrum in MG theories with primordial NG and show how both these deviations from standard case affect the observables. Then we define the galaxy growth rate parameter. In Sec. (\ref{Sec-Conc}) we conclude  and discuss the future prospects of this work. In this work we use the cosmological parameters  from recent Planck data as $\Omega_m=0.32$, $\Omega_{\Lambda}=0.68$, $n_s=0.96$ and $\ln(10^{10}A_{s})=3.1$ \cite{Ade:2013zuv}.


\section{Modified Growth rate}
\label{Sec-Mg}
One of the probable solutions of the accelerated expansion of the Universe is the modification of gravity in cosmological scales. There is a large amount of literature on different modified gravity models which is supposed to produce the late time acceleration of the  universe, like $f(R)$ theories of gravity in metric formalism \cite{Carroll:2003wy,Hu:2007nk}, in Palatini formalism \cite{Olmo:2011uz}, brane world models of Modified gravity \cite{Dvali:2000hr}, massive gravity \cite{D'Amico:2011jj} and etc.
In this work we examine the $f(R)$ gravity in metric formalism as a candidate for the accelerated expansion of the Universe and a parameterized model to show the deviations from $\Lambda$CDM cosmology,  in order to study the simultaneous effect of primordial NG and modified growth rate. In the first subsection we consider the Hu-Sawicki \cite{Hu:2007nk} model. In the second subsection we introduce the BZ parametrization as an alternative representation of MG theories and its relation to HS $f(R)$ model is discussed. In next sections we use the Hu-Sawicki $f(R)$ model.
\subsection{Hu-Sawicki $f(R)$ model}
One of the main concerns about $f(R)$ modifications of gravity is the solar system gravity test \cite{Erickcek:2006vf}. Khoury and Weltman \cite{Khoury:2003aq} propose a screening mechanism to evade solar system test. In this chapter we will use a $f(R)$ model that is viable and evade solar system tests.

 The $f(R)$ action of modified gravity is written as:
\be  \label{eq-EH}
S=\frac{1}{2\kappa^2}\int d^4x\sqrt{-g}f(R)+S_m(g_{\mu\nu},\chi_m)
\ee
 where $f(R)$ is a function of Ricci scalar, $\kappa^2\equiv 8\pi G $, and $\chi_m$ represents the matter fields. In the case of $f(R)=R$ we get the Einstein-Hilbert(EH) action. The deviation from EH action in cosmological scales is the cause of the cosmic expansion. In order to see this, we have to rewrite the modified Einstein field equations.  The corresponding modified Friedman equations are obtained from the variation of action in Eq.(\ref{eq-EH}) with respect to metric as \cite{Tsujikawa:2009ku}:
 \begin{eqnarray} \label{Eq-MFried}
 3FH^2&=&8\pi G\rho_m +\frac{FR-f}{2}-3H\dot{F},\\ \nonumber
 -2F\dot{H}&=&8\pi G\rho_m+\ddot{F}-H\dot{F}
 \end{eqnarray}
where $\rho_m$ is the matter density, $F\equiv\partial f/\partial R$ is the first derivative of the action with respect to Ricci scaler, which represent the degree of freedom of the action in the equivalent scalar tensor theory of the $f(R)$ \cite{Tsujikawa:2008uc} and finally $H=\dot{a}/a$ is the Hubble parameter. The Hubble parameter is related to the Ricci scaler as $R=6(2H^2 + \dot{H})$  where dot represents the derivative with respect to cosmic time hereafter. The expansion rate obtained from Eqs.(\ref{Eq-MFried}) for viable $f(R)$ theories is very close to the expansion rate obtained from $\Lambda$CDM \cite{Hu:2007pj}. In order to quantify the deviation from $\Lambda$CDM we define a dimensionless parameter $m$ as\cite{Tsujikawa:2009ku} :
\be
m=\frac{R F_{,R}}{F}
\ee
where $F_{,R}\equiv\partial F/\partial R$, and the  Ricci scaler is controlled by modified Hubble expansion rate. In the case of ($m=0$) we will recover the $\Lambda$CDM  universe.
However, the small deviation $m\ll 1$ is difficult to detect in the expansion history of the Universe. Also the MG theories in the background level are indistinguishable from smooth Dark Energy models \cite{Baghram:2010mc}. Consequently in studying the accelerated expanding universe models, we are interested in  the behavior of cosmological models in perturbation theory and LSS. The main point here is that the growth of structures in the Universe is a very promising tool to distinguish the MG models from DE/CC. This happens by introducing a scale dependent growth of perturbations.
 In order to study the LSS in MG theories, we need to use the perturbation theory. In Fourier space the density perturbation is obtained as \cite{Song:2006ej} :

\begin{eqnarray}\label{Eq-pert}
\ddot{\delta}_m+\left(2H+\frac{\dot{F}}{2F}\right)\dot{\delta}_m-\frac{8\pi G\rho_m}{2F}\delta_m&=&\frac{1}{2F}\left[(-6H^2+\frac{k^2}{a^2})\delta F+3H\dot{\delta F}+3\ddot{\delta F}\right], \\ \nonumber
\ddot{\delta F}+3H\dot{\delta F}+\left(\frac{k^2}{a^2}+\frac{f_{,R}}{3f_{,RR}}-\frac{R}{3}\right){\delta F}&=&\frac{8\pi G\rho_m\delta_m}{3}+\dot{F}\dot{\delta}_m
\end{eqnarray}

Considering the fact that the expansion history indicates that the time derivative of $F$ is small and also that $\ddot{\delta F}\ll H\dot{\delta F}\ll H^2$; the second order differential equation for evolution of density contrast by combining Eqs.(\ref{Eq-pert}) is  \cite{Tsujikawa:2008uc}:

\be \label{eq-delta}
\ddot{\delta}_m+2H\dot{\delta}_m-4\pi G_{eff}\rho_m\delta_m=0
\ee
where $G_{eff}$ is defined as:
\be \label{Eq-geff}
G_{eff}\equiv\frac{G}{F}\left[1+\frac{1}{3}\left(\frac{k^2}{a^2M^2/F+k^2}\right)\right]
\ee
in which $M^2\equiv R/3m = {F}/{3F_{,R}}$ is the effective mass of the scalaron, corresponding to the scaler field \cite{Tsujikawa:2008uc}. It is obvious that the scale dependence on the evolution of matter density is introduced via effective Newtonian constant in Eq.(\ref{Eq-geff}).
Now we are ready to study the deviation  from $\Lambda$CDM due to the MG theory.
The first parameter is the Growth function defined as $\delta_m(k,z)=D(k,z)\delta_m^i$, where $\delta_m^i$ is the initial value of the matter density contrast fixed in an initial redshift. (We can fix the growth function in present time as well.)
The other useful quantity is the logarithmic rate of change of matter density with respect to the  scale factor, known as growth rate.
The growth rate is defined as:
\be
f(k,z)=\frac{d\ln\delta (k,z)}{d\ln a}
\ee
where growth function now depends on wavenumber as well.
Now by using $\dot{\delta}=Hf\delta$ and $\ddot{\delta}=(H^2f^2+\dot{H}f+H\dot{f})\delta$, we can rewrite Eq.(\ref{eq-delta}) in terms of growth rate and its derivative as:
\be \label{eq-f}
\dot{f}+Hf^2+\left(2H+\frac{\dot{H}}{H}\right)f -\frac{3}{2}\Omega_m\frac{G_{eff}}{G}\frac{H^2_0}{H}=0
\ee
 where the growth rate  is related to the scalaron mass $M$, the action derivative $F$, the expansion rate of cosmos $H$ and the wavenumber that we observe the structures $k$ via Eq.(\ref{Eq-geff}). Now we write the  Eq.(\ref{eq-f}) in terms of redshift:
\be \label{Eq-f}
f'(k,z)-\frac{f^2(k,z)}{1+z}+\left(\frac{E'(z)}{E(z)}-\frac{2}{1+z}\right)f(k,z)+\frac{3}{2}\Omega^0_m\frac{(1+z)^2}{E^2(z)}\frac{G_{eff}(k,z)}{G}=0
\ee
where $E(z)\equiv H(z)/H_0$, and $\Omega^0_m$ is the present value of matter density. Prime denotes derivative with respect to redshift. Now  by solving Eq.(\ref{Eq-MFried}) and Eq.(\ref{Eq-f}), we can find the growth rate of the structures which depends on the wavenumber via the $G_{eff}$ term \cite{Baghram:2010mc}.
One type of MG parametrization is to use a free parameter $\gamma$ to parameterize the growth rate $f=\Omega(z)^{\gamma}$. In table \ref{table-b}, we report the constrains on free parameter $\gamma$.

\begin{figure}[t]
\centering
\includegraphics[width=10cm]{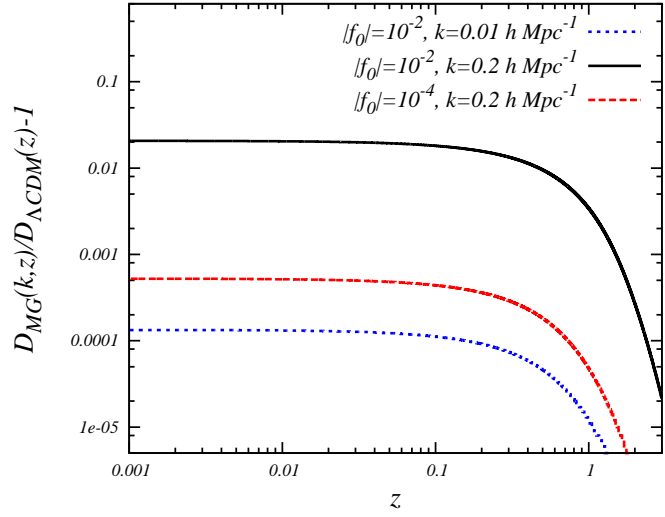}
\caption {The ratio of growth function $D_{MG}(k,z)/D_{\Lambda CDM}-1$ for MG, HS model with $|f_0|=10^{-2}$, $k=0.2 h/Mpc$ (black solid line) and $\Lambda CDM$ is plotted versus redshift. The red-long dashed line indicate the ratio for $|f_0|=10^{-4}$, $k=0.2 h/Mpc$ and the MG model with $|f_0|=10^{-2}$, $k=0.01 h/Mpc$ is plotted in blue dotted line.} \label{Fig-D}
\end{figure}

 In order to be more precise and discuss the effect of growth rate on the  redshift distortion and the galaxy power spectrum  in Sec.(\ref{Sec-galaxy}) we choose a specific model of MG. The  Hu and Sawicki \cite{Hu:2007nk} model
\be\label{Eq-hu}
f(R)=R-\mu R_c\frac{(R/R_c)^2}{1+(R/R_c)^{2}}
\ee
in which $\mu>0$ and $R_c>0$ are free parameters. This model evades solar system constraints of MG. In order to quantify the deviation from $\Lambda$CDM with only one free parameter, we write the first derivative of the action as $F\equiv 1+\tilde{F}$ by assuming that the action is $f(R)=R+\tilde{f}(R)$ and $\tilde{F}\equiv\partial\tilde{f}/\partial R$.
For the action introduced in Eq.(\ref{Eq-hu}) we will have:
\be
\tilde{F}=-2f_0 \frac{R}{H^2_0}\left[1+(\frac{R}{R_c})^2\right]^{-2}
\ee
where $|f_0|\equiv (\mu H_0^2)/{R_c}$ is the free parameter of the model, $H_0$ is the present value of Hubble parameter, and $R_c$ is assumed to be the present value of the Ricci scalar.
In Table (\ref{table-b}), we summarize the observational constraints on HS Model obtained from different geometrical and dynamical observations. We also mention the different conventions used to parameterize the degrees of freedom of the model.
\begin{table}[ht]
\caption{Growth rate $f_{obs}$ observational constraints}
\centering
\begin{tabular}{c c c c c}
\hline\hline
$Parameter$ & $ Observations$ & $Constraints$ & Note & Ref. \\ [0.5ex]
\hline
$|f_0|$ &  SNIa \cite{Kowalski:2008ez} , BAO \cite{Eisenstein2006}, age \cite{Simon:2004tf}   & $<0.03 (1\sigma)$r
& one more free parameter &  \cite{Martinelli:2009ek} \\
$B_0$ &  cluster abundance + SNI + BAO + gISW &  $1.1\times 10^{-3} (2\sigma)$  &  $B_0=\frac{F,R}{F}\frac{dR}{dz}\frac{H}{dH/dz}$  &    \cite{Lombriser:2010mp} \\
$B_0 $&  CMB+BAO+ $H_0$ (HST)+ SNIa(Union2.1) &  $0.079 (1\sigma)$  &  BZ parametrization \cite{Bertschinger:2008zb} &    \cite{Hu:2013aqa}\\
$B_0$ &  CMB (WMAP 9y) + lensing & $0.473 (1\sigma)$  &  BZ parametrization \cite{Bertschinger:2008zb} &    \cite{Hu:2013aqa}\\
$B_0$ &  CMB(Planck) + Polarization (WMAP) & $0.849 (1\sigma)$  &  BZ parametrization \cite{Bertschinger:2008zb} &    \cite{Hu:2013aqa}\\
 $\gamma$ &  CMB power spectrum+ CMB bispectrum & $0.555^{+0.034}_{-0.042} (2\sigma$)&  $f=\Omega^\gamma(z)$ &  \cite{DiValentino:2012yg}  \\
$\mu$ &  galaxy growth rate $f\sigma_8$ &  $\mu>12$ &  HS model with one free parameter $n=1.5$  &  \cite{Okada:2012mn} \\
\hline
\end{tabular}
\label{table-b}
\end{table}

Now using the derivative of $\tilde{F}$ with respect to Ricci and substituting in the definition of the effective mass $M$ and accordingly in $G_{eff}$, we can solve the differential Eq.(\ref{eq-delta}) and Eq.(\ref{Eq-f}) for different wavenumbers with respect to redshift in order to find the growth function $D(k,z)$ and growth rate $f(k,z)$ respectively.
The $M$ parameter for HS model, is defined as:
\be
M^2=\frac{1-2f_0\tilde{R}\left[1+({\tilde{R}}/{\tilde{R}_c})^2\right]^{-2}}{2f_0[1+({\tilde{R}}/{\tilde{R}_c})^2]^{-3}\times[3({\tilde{R}}/{\tilde{R}_c})^2-1]}
\ee
where $\tilde{R}=R/H_0^2$ and $\tilde{R}_c=R_c/H_0^2$
\begin{figure}[t]
\centering
\includegraphics[width=10cm]{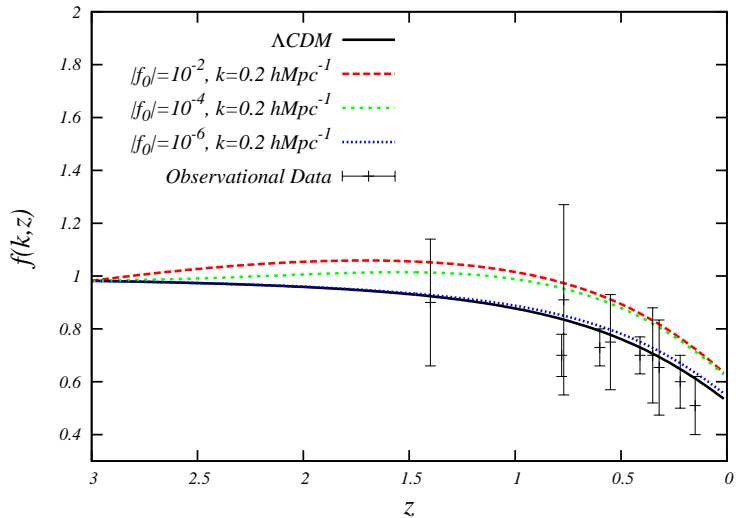}
\caption {The growth rate versus redshift is plotted for $k=0.2 h/Mpc$ for $|f_0|=10^{-2},10^{-4},10^{-6}$  with red long dashed, green dashed and blue dotted lines, respectively. For comparison we also plot the $\Lambda$CDM growth rate (solid black curve). The data points are taken from LSS surveys listed in Table (\ref{table-f}) with $1\sigma$ error-bars.} \label{Fig-fzkMG}
\end{figure}

\begin{table}[ht]
\caption{Growth rate $f_{obs}$ observational constraints}
\centering
\begin{tabular}{c c c c c}
\hline\hline
$z$ & $f_{obs}$ & $\sigma$ & Survey & Ref. \\ [0.5ex]
\hline
0.15 & 0.51 & 0.11 & 2dF &  \cite{Hawkins:2002sg,Linder:2007nu, Verde:2001sf} \\

0.22 & 0.60 & 0.10 & WiggleZ &     \cite{Blake:2011rj} \\

0.32 & 0.654 & 0.18 & SDSS &    \cite{Reyes:2010tr} \\

0.35 & 0.70 & 0.18 & SDSS &    \cite{Tegmark:2006az} \\

0.41 & 0.70 & 0.07 & SDSS &    \cite{Blake:2011rj}  \\

0.55 & 0.75 & 0.18 & 2dF-SDSS &    \cite{Ross:2006me} \\

0.60 & 0.73 & 0.07 & SDSS &  \cite{Blake:2011rj} \\

0.77 & 0.91 & 0.36 & VIMOS-VLT & \cite{Guzzo:2008ac} \\

0.78 & 0.70 & 0.08 & SDSS &  \cite{Blake:2011rj}\\

1.4 & 0.90 & 0.24 & 2dF-SDSS &  \cite{daAngela:2006mf} \\

3.0 & 1.46 & 0.29 & SDSS &  \cite{McDonald:2004xn} \\ [1ex]
\hline
\end{tabular}
\label{table-f}
\end{table}
In Fig.(\ref{Fig-D}), we plot the ratio of growth function $D_{MG}(k,z)/D_{\Lambda CDM}-1$ for MG, HS model with $|f_0|=10^{-2}$, $k=0.2 h/Mpc$ (black solid line) and $\Lambda CDM$ is plotted versus redshift. The red-long dashed line indicate the ratio for $|f_0|=10^{-4}$, $k=0.2 h/Mpc$ and the MG model with $|f_0|=10^{-2}$, $k=0.01 h/Mpc$ is plotted in blue dotted line. Fig.(\ref{Fig-D}) shows that the deviation from $\Lambda CDM$ is larger in small scales.
In Fig.(\ref{Fig-fzkMG}), we plot the growth rate versus redshift for a specific wavenumber $k=0.2 h/Mpc$ and different values of free parameter $|f_0|$.

The viable $f(R)$ gravity models have two regimes: I) $a^2M^2\ll k^2$, we will be in scalar tensor mode where the effective gravitational constant reaches  $4/3 G$;  II) in the regime of very massive scalaron $k^2\ll a^2M^2$, we will recover the $\Lambda$CDM case. It is obvious from Fig.(\ref{Fig-fzkMG}) that for all values of $f_0$, the growth rate is higher than the $\Lambda$CDM case. This is because of the enhancement of the gravitational constant in all cases with an enhancement factor $G_{eff}/G$ running from $1$ to $4/3$. In Fig.(\ref{Fig-fzkMG}), it is shown that  by increasing the deviation parameter from CC  $|f_0|$,  the growth rate deviates more from CC case. The observational data points are from different surveys with $1\sigma$ error-bar listed in Table (\ref{table-f}).  Fig.(\ref{Fig-fzkMG}) shows that the recent growth rate measurement can not rule out the MG with small deviations $f_0\leq 10^{-5}$. The future observations like LSST  \cite{Zhan:2006gi} and Euclid \cite{Amendola:2012ys}  can constrain the growth rate at intermediate redshifts. Another important point to indicate is that in the wavenumber $k\simeq aM$(which depends on the model parameters, the growth of the structures changes its regime, where we can see this effect in matter power spectrum in upcoming sections.

\begin{figure}[t]
\centering
\includegraphics[width=10cm]{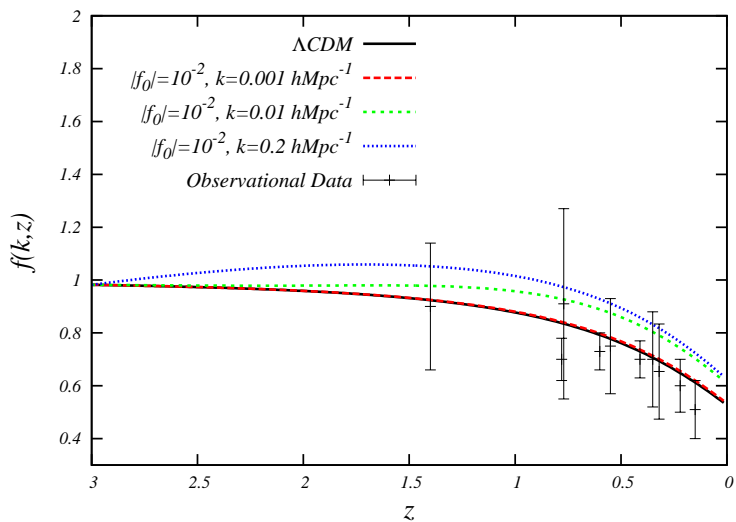}
\caption {In this figure we plot the growth rate versus redshift with $|f_0|=10^{-2}$ for wavenumbers $k=0.001,0.01,0.2 ~h/Mpc$  with blue dotted, green dashed and red long dashed lines, respectively. For comparison we also plot the $\Lambda$CDM growth rate (solid black curve). The data points are taken from LSS surveys listed in Table (\ref{table-f}) with $1\sigma$ error-bars. } \label{Fig-fzk}
\end{figure}

In Fig.(\ref{Fig-fzk}), we study the scale dependence of the growth rate in MG. We set the free parameter $|f_0|=10^{-2}$ and plot the growth rate versus the redshift for wavenumbers $k=0.01,0.1,0.2 h/Mpc$, respectively. As discussed previously for very small wavenumbers (large structures), the $f(R)$ gravity is completely indistinguishable from $\Lambda$CDM. The main effect of these theories are in quasi-linear regimes, large wavenumbers (smaller structures) where the growth rate has a strong scale dependence and deviates from the standard case. The observational data points are taken from Table(\ref{table-f}). In this subsection, we have discussed the scale dependence growth rate in HS model. In the upcoming subsection we introduce the  BZ parametrization and its relation to HS model.

\subsection{Bertschinger-Zukin parametrization}

In this subsection, we introduce the BZ parametrization and its relation with HS model.
In general, the chameleon, symmetron and dilaton screening MG theories can be written in Einstein frame as an Einstein Hilbert action plus a new degree of freedom, which is coupled to dark matter/baryons via conformal factor:
\be
S_{E}=\int d^4x\sqrt{-\tilde{g}}\left[\frac{M^2_{pl}}{2}\tilde{R} - \frac{1}{2}\tilde{g}^{\mu\nu}(\tilde{\nabla _{\mu}\phi})(\tilde{\nabla}_{\nu}\phi)-V(\phi)\right]+S_{i}[\chi_i, e^{-\kappa\alpha_i(\phi)}\tilde{g}_{\mu\nu}]
\ee
where $\chi_i$ represent the matter fields and the Jordan frame metric is related to the Einstein frame metric by  conformal factor:
\be
g_{\mu\nu}=e^{-\kappa\alpha_i(\phi)}\tilde{g}_{\mu\nu}
\ee
now in order to study the effect of MG on linear perturbations Bertschinger and Zukin \cite{Bertschinger:2008zb} define the scale dependent effective Newtonian constant $G\mu(k,z)$ and gravitational slip parameter $\gamma(k,z)$ as:
\be
k^2\Psi = -4\pi Ga^2\mu(k,z)\rho_m\delta_m
\ee
\be
\frac{\Phi}{\Psi}=\gamma (k,z)
\ee
where $\Psi$ and $\Phi$ are metric perturbation terms in perturbed FRW ( $ds^2=-(1+2\Psi)dt^2+a^2(1-2\Phi)dx^idx_i$).
Now by using the fact that the deviation from $\Lambda$CDM model, can be parameterized by one parameter \cite{Song:2006ej}, Bertschinger and Zukin parameterize the deviation parameters as:
\be
\mu(k,z)=\frac{1+{\frac{4}{3}\lambda^2 k^2}/{(1+z)^4}}{1+{\lambda ^2k^2}/{(1+z)^4}},
\ee
\be
\gamma (k,z)=\frac{1+{\frac{2}{3}\lambda ^2 k^2}/{(1+z)^4}}{1+{\frac{4}{3}\lambda^2 k^2}/{(1+z)^4}}
\ee
where the free parameter is the compton-wavelength of new degree of freedom. $\lambda$ is related to the free parameter $B_0$ as:
\be
\lambda ^2 \equiv \frac{B_0}{2H_0^2}
\ee
In table \ref{table-b}, we report some observational constraints on $B_0$. Now the $B_0$ parameter is related  to the derivatives of $f(R)$ as:
\be
B=\frac{F_{,R}}{F}R\frac{H}{H'}
\ee
where $R$ is the Ricci scalar, which can be assumed to the has the same value as $\Lambda$CDM, because the viable $f(R)$ models can mimic the background evolution of $\Lambda$CDM. In this case we can relate the free parameter of HS model $f_0$ to the model independent free parameter of HZ parametrization as:
\be
|f_0|\simeq|\frac{1}{B_0}\frac{R(z=0)H_0}{H'(z=0)}-1|
\ee
In the next section, we will investigate the effect of primordial NG on bias parameter and then we will study the simultaneous effect of modified gravity growth and scale dependence bias in special case of HS model. The general study of this effect on MG models is out of the scope of this work.
\section{Non-Gaussian Bias}
\label{Sec-bias}
In this section we will discuss the bias parameter and its application to study the effect of the primordial NG and MG on it. As mentioned in the introduction, the statistics of the luminous matter in the Universe is a promising tool to constrain the cosmological models. To study their statistics we need to know about the structure formation in nonlinear regime and the relation between the dark matter halos and luminous matter \cite{Mo:1995cs}.
The structures in the Universe (i.e galaxies and cluster of galaxies) are hosted by  dark matter halos. These dark matter halos are formed, when a region of a mass $M$, with a density contrast of $\delta =(\rho-\bar{\rho})/\bar{\rho}$, exceeds the critical density threshold. (for spherical collapse $\delta_c\simeq 1.68$)\cite{Gunn:1972sv}.
The abundance of dark matter halos are modulated by the background density of matter perturbations. This modulation is formalized by the halo bias parameter \cite{Bardeen:1985tr}.
In this context the bias parameter is defined as the ratio of the the halo density  contrast $\delta_h$ and the background matter density $\delta_m$ (In this work we assume that the galaxy-halo bias is unity $\delta_g=\delta_h$):
\be
b=\frac{\delta_h}{\delta_m}
\ee
The bias parameter can be obtained in the framework of Peak-Background Splitting \cite{Sheth:1999mn}. In that case the bias can be obtained from the probability of the structures formed above a threshold height $\nu\equiv \delta_c / \sigma(M)$ where $\sigma(M)$ is the variance of density contrast. In this picture the bias is defined as:
\be \label{eq-bl}
b_L= - \frac{1}{\sigma(M)}\frac{d\ln \bar{n}(M,\nu)}{d \nu}
\ee
where $b_L$ is the Lagrangian bias and $\bar{n}$ is the mean number density of the structures with mass $M$ and height $\nu$. In the Eq.(\ref{eq-bl}) we assume that we can split the matter density as $\delta_h=\delta_s+\delta_l$ where $\delta_s$ indicate the short wavelength density contrast of matter corresponding to the structures and $\delta_l$ is the long wavelength mode with the condition $\delta_l\ll\delta_s$.
If we assume the very simple case of Press-Schechter mass function as:
\be
\bar{n}(M,z)=-2\frac{\bar{\rho}}{M^2}f(\nu)\frac{d\ln\sigma(M,z)}{d\ln M}
\ee
with
\be
f(\nu)=\frac{\nu}{\sqrt{2\pi}}e^{-\nu^2/2}
\ee
The Eulerian bias parameter can be written:
\be
b_{E}(z)=1+\frac{\nu^2(z)-1}{\delta_c}
\ee

 Using the Press-Schechter mass function \cite{Press:1973iz} and with  the Gaussian initial function, the bias parameter becomes scale independent. This is almost true for all Gaussian initial conditions with the assumption of the universality of mass function, which means the probability of finding structure in a mass range is only dependent to the height parameter.
Recently, it was shown that the primordial NG will introduce a scale dependent bias \cite{Dalal:2007cu}. After that
a huge amount of literature is devoted to the study of the effect of NG in LSS observations specially the bias parameter. \cite{LoVerde:2007ri,Matarrese:2008nc, Afshordi:2008ru,Jeong:2009vd,Verde:2010wp,Desjacques:2010nn,Norena:2012yi}

 The idea is that in the local NG, the background perturbation is modulated by the NG case as:
\be
\Phi_{NG}=\Phi_G+f_{NL}\Phi^2_{G}
\ee
where $\Phi$ is the Bardeen potential which can be used instead of the density perturbation to study the effect of NG. It is obvious that we can use Bardeen potential and matter density contrast interchangeably by using the Poisson equation as a link.
Dalal et al. \cite{Press:1973iz} showed that in the case of local non-Gaussianity the bias parameter is:

\be\label{Eq-bNG}
b_{NG}=\frac{2f_{NL}(b_E-1)\delta_c}{{\cal{M}}(k,z)}
\ee
 where the ${\cal{M}}$ is a function relating the primordial curvature perturbation ${\cal{R}}$ to the linear density functions $\delta_m$  as ${\cal{M}}=\delta_m/{\cal{R}}$. The form of {\cal{M}} is as follows:
\be
{\cal{M}}=\frac{2}{5}\frac{k^2 T(k)D(z)}{H^2_0\Omega^0_m}
\ee
where $D(z)$ is the growth function and $T(k)$ is the transfer function.
It is worth to note that the Eq.(\ref{Eq-bNG}) can be obtained, from the Excursion Set Theory approach \cite{Bond:1990iw}. This is discussed  in App.(1) of \cite{Baghram:2013lxa}
In this section, the linear Gaussian and local NG bias for $\Lambda$CDM and MG theories are calculated.
For this task, we use the mass function of Sheth-Tormen \cite{Sheth:1999mn} to find the linear bias.
The Sheth-Tormen probability function $f(\nu)$ is defined as:
\be
f_{ST}=A\sqrt{\frac{\alpha\nu^2}{2\pi}}\left[1+\frac{1}{(\alpha\nu^2)^p}\right]e^{-\frac{\alpha\nu^2}{2}}
\ee
where $A\simeq0.32$, $\alpha\simeq 0.707$ and $p\simeq 0.3$ are free parameters obtained from N-body simulation \cite{Sheth:1999mn}.
The Sheth-Tormen linear bias will be:
\be
b^{L}_{ST}(M,z)=1+\frac{1}{\delta_c}\left[\alpha\nu^2(z)-1+\frac{2p}{1+(\alpha\nu^2(z))^p}\right]
\ee
In order to calculate the NG  bias defined in Eq.(\ref{Eq-bNG}) for the  $\Lambda$CDM model, we use the standard growth function and the  Bardeen, Bond, Kaiser and Szalay(BBKS) transfer function \cite{Bardeen:1985tr} defined  respectively as follows:
\begin{equation}
D(z)=\frac{5}{2}\frac{1}{1+z}\Omega_{m}\left[\Omega^{4/7}_{m}-\Omega_{\Lambda}+(1+\frac{\Omega_m}{2})(1+\frac{\Omega_{\Lambda}}{70})\right]^{-1}
\end{equation}
and
\begin{equation}
T(k=q\Omega_m h^2 Mpc^{-1})\approx \frac{\ln[1+2.34q]}{2.34q}\times\left[1+3.89q+(16.2q)^2+(5.47q)^3+(6.71q)^4\right]^{-1/4} 
\end{equation}
where $\Omega_m=\Omega^0_m a^{-3}/(\Omega_m a^{-3}+\Omega_{\Lambda})$ and  $\Omega_{\Lambda}=\Omega^0_{\Lambda}/(\Omega_m a^{-3}+\Omega_{\Lambda})$.

\begin{figure}[t]
\centering
\includegraphics[width=9cm]{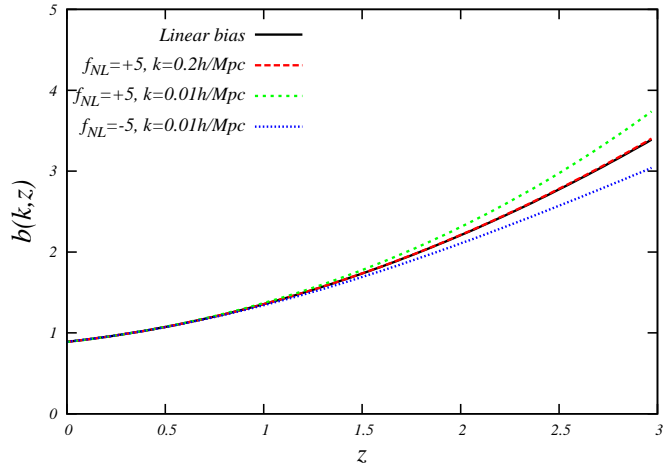}
\caption {The bias parameter versus redshift is plotted for $f_{NL}=+5$ and $k=0.2 h/Mpc$ (red long dashed line), $f_{NL}=+5$ and $k=0.01 h/Mpc$ (green dashed line) and $f_{NL}=-5$ and $k=0.01 h/Mpc$ (blue dotted line). For comparison, we also plot the Gaussian Sheth-Tormen linear bias with solid black line.} \label{Fig-bias}
\end{figure}

In Fig.(\ref{Fig-bias}), we plot the bias parameter versus redshift for the linear case using the Sheth-Tormen mass function. We also plot the local non-Gaussian bias with $f_{NL}=\pm 5$, for different wavenumbers. As we expect, there is a $k^{-2}$ dependence in bias. The largest deviation appears for small wave-numbers.
A positive $f_{NL}$ increases the bias parameter whereas a negative $f_{NL}$ decreases the bias in comparison to the linear case.
The NG bias for $k=0.2 h/Mpc$ is almost indistinguishable from the $\Lambda$CDM case. The bias parameter measurement now are done in redshift averaged bins which are not as accurate to distinguish between the cosmological models with small local non-Gaussianity\cite{Blake:2011rj}. Although the future observations are very promising to reach the precision of Planck satellite to detect the local NG \cite{Verde:2010wp}.

\begin{figure}[t]
\centering
\includegraphics[width=10cm]{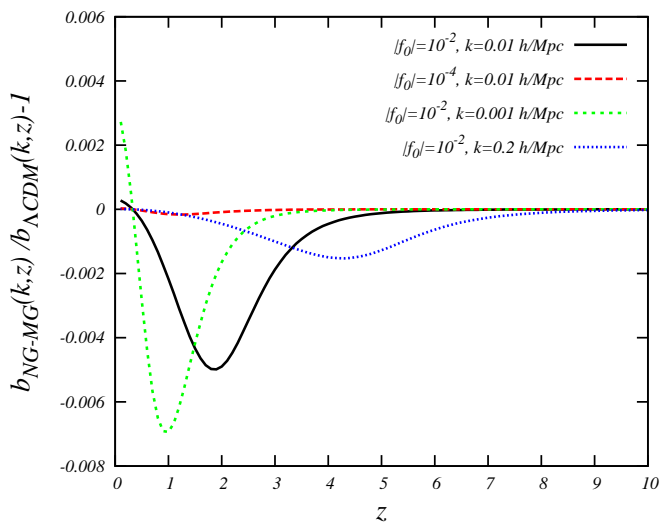}
\caption {The NG bias ratio of $b^{MG}/b_{\Lambda CDM}-1$ is plotted versus redshift. The red long dashed line is for deviation parameter of $|f_0|=10^{-4}$ with wavenumber $k=0.01 h/Mpc$. For deviation of $|f_0|=10^{-2}$ we plot the bias ratio for the wavenumbers $k=0.001 , 0.01 , 0.2 h/Mpc$ with green dashed line, solid black line and blue dotted line respectively.} \label{Fig-biasMG}
\end{figure}

In the case of the MG, the non-Gaussian bias is also modified through the Poisson equation. Assuming that the universality of the mass function is unchanged in the MG, the only effect is imprinted in the relation between the curvature perturbation and matter density through the modified ${\cal{M}}_{MG}$:
\be \label{Eq-Meff}
{\cal{M}}_{MG}= \frac{2}{5}\frac{G_{eff}}{G}\frac{k^2 T(k)D_{MG}(z,k)}{H^2_0\Omega^0_m}
\ee
where $D_{MG}$ is MG growth function and $G_{eff}/G$ is the gravitational constant enhancement discussed in Sec.(\ref{Sec-Mg}).  The modified growth function, can be obtained from the  relation $\dot{D}_{MG}(k,z)=H f_{MG}(k,z)D_{MG}(k,z)$  with the same initial conditions as $\Lambda$CDM in dark matter dominated era. The  modified growth rate can be obtained by solving Eq.(\ref{Eq-f}).

 A crucial point is the appearance of a new scale dependence in the NG bias parameter via the modified growth function and also an effective Newtonian constant in Eq.(\ref{Eq-Meff}). An important point to indicate is that this modification is almost concrete for the short wavenumbers (local case). The study of the general case for desired bispectrum is out of the scope of this work.

 In Fig.(\ref{Fig-biasMG}), we plot the  ratio of the total bias (Gaussian+ NG)
   in MG Modified gravity introduced in previous section to the $\Lambda$CDM versus redshift.  We plot this ratio for the deviation of MG with $|f_0|=10^{-4}$ with wavenumber $k=0.01 h/Mpc$ and also a larger deviation with $|f_0|=10^{-2}$ for two wavenumbers $k=0.01 h/Mpc$ and $k=0.2h/Mpc$ respectively.
The NG bias with MG is the same as the $\Lambda$CDM, non-Gaussian case for large wavenumbers, this is because the NG bias is very small and indistinguishable with linear bias for large structures. Instead for small wavenumbers where local NG effect is most efficient a new scale dependence is introduced by MG. The future very large scale surveys are promising to constraint the Modified gravity NG bias parameter. An important point is that this ratio goes to zero in higher redshifts. This because the scale dependence introduced by MG is vanished in higher redshifts.

In the next section we will discuss the redshift space distortion parameter $\beta=f/b$ and its scale dependence introduced by NG and MG. Then we will discuss the RSD effect on the galaxy power spectrum.

\section{Large Scale Structure Observables}
\label{Sec-galaxy}

In the previous sections we showed that the growth rate and the bias parameter are both modified by the primordial NG and the deviation from $\Lambda$CDM. In this section, in first subsection we investigate the effect of these two parameters on Redshift Space Distortion (RSD) and galaxy power spectrum. In the second subsection we introduce the galaxy growth rate  and discuss about its relation with the matter growth rate.

 \subsection{Redshift space distortion and galaxy power spectrum}
 The idea of the redshift space, as explained in introduction, is that what we observe is the luminous matter distribution in redshift space instead of the spatial coordinate. The clustering of the matter changes the peculiar velocity of the dark matter tracers and affect the observed redshift. Accordingly by finding the amount of distortion occurred from transferring the redshift coordinate to real coordinate will be a measure for the amount of clustered matter.
The  density contrast of matter in redshift space is related to the real space via:

\be
\delta_z(k)=\delta_r(k)\left(1+\beta\mu^2\right)
\ee
where $\mu={\bf{k}}.{\bf{r}}/kr$ is the direction between the wavenumber and the line of sight and $\beta$ is the redshift space parameter defined as:
\be
\beta(k,z)=\frac{f(k,z)}{b(k,z)}
\ee
where the $\beta$-function in contrast to the $\Lambda$CDM case is a scale dependent parameter. This scale dependent behavior is introduced by the NG bias on one hand and the MG growth rate function on the other hand.

\begin{figure}[t]
\centering
\includegraphics[width=10cm]{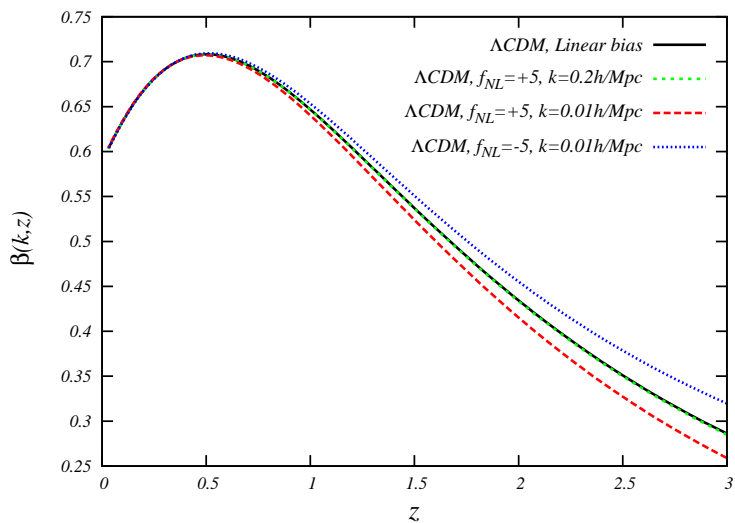}
\caption {The redshift space distortion parameter $\beta=f/b$ is plotted versus redshift. The solid black line shows the $\Lambda$CDM growth rate with Sheth-Tormen linear bias. The green dashed line shows the $\beta$ with the non-Gaussian bias with $f_{NL}=+5$ and the wavenumber $k=0.2$. The long dashed red line is with the non-Gaussian bias with $f_{NL}=+5$ with wavenumber $k=0.01$ and the blue dotted line is the $\beta$ function for $f_{NL}=-5$ and the wavenumber $k=0.01$. In all NG cases we use the $\Lambda$CDM growth rate.} \label{Fig-beta}
\end{figure}

In Fig.(\ref{Fig-beta}), we plot the $\beta$-function for the non-Gaussian case. In the NG case, $\beta$ for wavenumber $k=0.2 h/Mpc$ is almost indistinguishable from the linear case. This is because the NG shows up in small wavenumbers. The important point here is that for positive local NG, say $f_{NL}=+5$, the NG bias is increased which causes a decrease in $\beta$ -function as it is proportional to the inverse of bias. The negative NG increase the redshift space distortion in higher redshifts in comparison with the linear case.
Now we want to explore the effect of the MG on the $\beta$-function.

\begin{figure}[t]
\centering
\includegraphics[width=10cm]{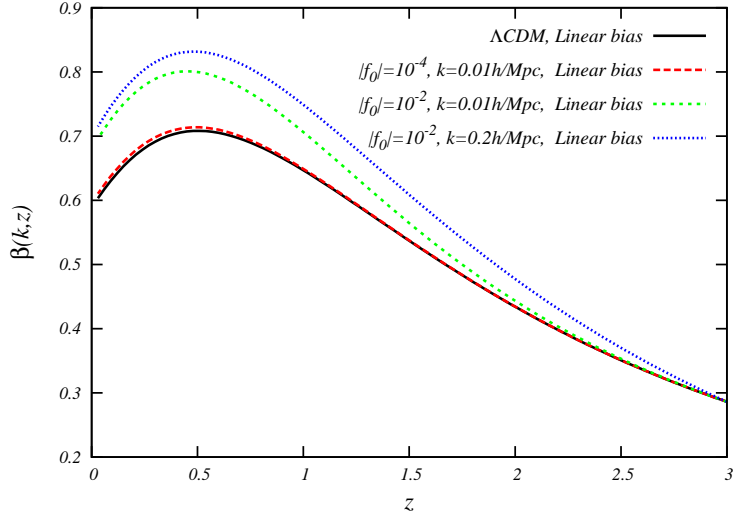}
\caption {The redshift space distortion parameter is plotted versus redshift for $f(R)$. The black solid line represents the $\Lambda$CDM RSD parameter with linear bias. The  red long dashed line represents the $|f_{0}|=10^{-4}$ with wavenumber $k=0.01 h/Mpc$. The green dashed line, shows the $|f_{0}|=10^{-2}$ with $k=0.01 h/Mpc$The blue dotted shows $|f_{0}|=10^{-2}$ with $k=0.2h/Mpc$. In all cases the bias parameter is linear.} \label{Fig-betaMG}
\end{figure}

In Fig.(\ref{Fig-betaMG}) we plot the RSD parameter versus redshift for MG case with the deviations of $|f_0|=10^{-4}$ and $|f_0|=10^{-2}$. The deviations from $\Lambda$CDM  case is almost negligible for $|f_{0}|=10^{-4}$. In the case of  larger deviation the scale dependence of the RSD is affected by MG. For large wavenumbers we have the largest deviation from the standard case. The main deviation occurs for larger wavenumbers, as we expected.

\begin{figure}[t]
\centering
\includegraphics[width=10cm]{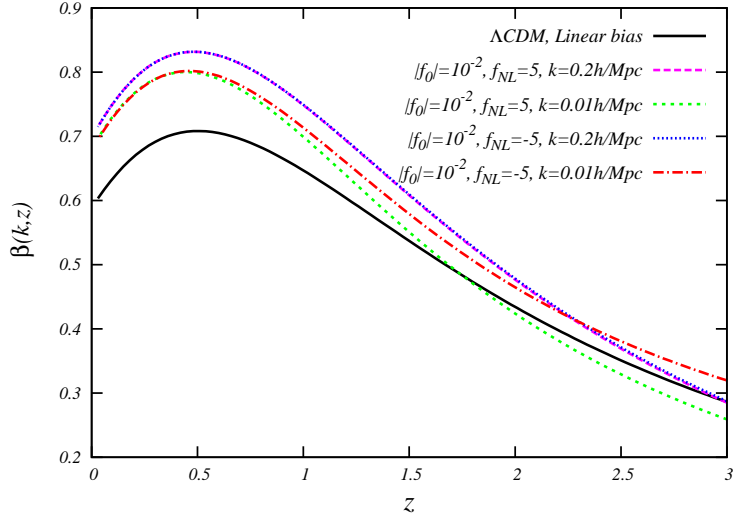}
\caption {The $\beta$-function is plotted versus redshift for $f_{NL}=+5$  with wavenumbers $k=0.2 h/Mpc$ and $k=0.01 h/Mpc$ with purple long dashed line and green dashed line. We also plot the RSD parameter for negative NG $f_{NL}=-5$ with wavenumbers $k=0.2 h/Mpc$ and $k=0.01 h/Mpc$ with blue dotted line and dash-dotted red line. For comparison we plot the linear $\Lambda$CDM case with solid black line.} \label{Fig-betaMGNG}
\end{figure}

Now the interesting point here is the study of the combination of the two  effects of NG and MG on $\beta$-function.
As the effect of NG is on small wavenumbers and the effect of MG on large wavenumbers, we will assume a non-trivial combination of the scale dependence.
In Fig. (\ref{Fig-betaMGNG}) we plot the $\beta$-function versus redshift for positive and negative local NG amplitude $f_{NL}=\pm 5$ and two wavenumbers $k=0.01, 0.2 h/Mpc$.  The green dashed line shows the $\beta$-function for MG with deviation $|f_0|=10^{-2}$ and a positive local NG $f_{NL}=+5$ with wavenumber $k=0.01 h/Mpc$. In small redshifts, the deviation from CC, has the dominate effect on the RSD parameter. Where in the higher redshifts the effect of positive NG dominates.
 For the wavenumber $k=0.2 h/Mpc$, the effect of NG is small so in higher redshifts the effect of positive and negative NG is indistinguishable. The largest $\beta$-function value in higher redshifts correspond to negative NG and small wavenumber $k=0.01 h/Mpc$ which cause to a decrease in bias parameter and correspondingly to higher $\beta$ function.

The other LSS observable is the power spectrum of galaxies.
Now we can transfer the RSD effect to the galaxy power spectrum  using the relation between the square of the real and redshift space density contrast  $P^{(z)}=P^{(r)}(1+\beta\mu^2)^2$ and averaging over the angle $\mu$
\be
P^{(z)}_g(k,z)= b^2 P^{(r)}_m(k,z)\left[1+\frac{2}{3}\beta +\frac{1}{5}\beta^2\right]
\ee
Where  $b$ is the total bias parameter , defined in Sec.(\ref{Sec-bias}) which is a sum of the linear and non-Gaussian terms $b=b_L+b_{NG}$ and $f$ is the growth rate of MG theory.
$P^{(r)}_m$ is the linear matter power spectrum related to the growth function $D(z)$
 and  the transfer function $T(k)$ as:
 \be
 P_m(k,z)=A k^{n_s}T^2(k)D^2(z)
 \ee
 where $A$ is the amplitude of matter fluctuations, $n_s$ is the spectral index, and the $T(k)$ and $D(z)$  are the transfer function and growth function, respectively. For the MG case we substitute $D(z)$ with $D_{MG}(z,k)$  and assume that the transfer function is the same for both theories. In Fig.(\ref{Fig-Pg}) we plot the galaxy power spectrum  versus redshift for $\Lambda$CDM case with linear bias (red-solid line) and for the modified gravity theory with the deviation parameter $|f_0|=10^{-2}$ and the local NG with $f_{NL}=+5$ (blue-dotted line) with non-linear corrections. The data points are the galaxy power spectrum from the Luminous Red Galaxy (LRG) data from SDSS survey \cite{Tegmark:2006az}.
As it was shown in Fig.(\ref{Fig-Pg}), the primordial NG effect with the MG background is the enhancement of the galaxy power spectrum. {To show the scale dependence of this enhancement, in Fig. (\ref{Fig-Pgrel}) we plot the ratio of the galaxy power spectrum for the modified gravity theory with the deviation parameter $|f_0|=10^{-2}$ and with local NG initial conditions with $f_{NL}=+5$  ($P_g^{MG-NG}$)and $\Lambda$CDM galaxy power spectrum with linear bias.
Accordingly this ratio is related to:
\begin{equation}\label{eq-ratio}
\frac{P_g^{NG-MG}}{P^{\Lambda CDM}}-1 =\frac{(b_L+b_{NG})^2}{b^2_L}\frac{D_{MG}(k,z)}{D(k,z)}\frac{1+\frac{2}{3}\tilde{\beta+}\frac{1}{5}\tilde{\beta}^2}{1+\frac{2}{3}\beta +\frac{1}{5}\beta ^2}-1
\end{equation}
where $\tilde{\beta}=f_{MG}/(b_L+b_{NG})$
Considering the fact that we are probing this ratio in low redshifts, where the contribution of terms with bias NG bias parameter is small due to Eq.(\ref{Eq-bNG}), where $b_{NG}\propto (1+z)$. Consequently Eq.(\ref{eq-ratio}) can be approximated by
\begin{equation}
\frac{P_g^{NG-MG}}{P^{\Lambda CDM}}-1 \simeq \left(1+2\frac{b^{NG}}{b_L}\right)\frac{D_{MG}(k,z)}{D(k,z)}-1
\end{equation}
 where we neglect the redshift space distortion term, because of Fig.(\ref{Fig-betaMGNG}) in low redshifts. The main contribution to the galaxy power spectrum comes from the modified gravity (in small scales) than the bias parameter. This is because we compare the two power spectrums in redshift range of $0.155 < z < 0.474$ of the LRG sample with median of $z\sim0.3$ which the nonlinear bias is small in comparison to linear bias as it was shown in Fig.(\ref{Fig-bias}). Consequently the main contribution comes from modified growth function which deviates from standard case in small scales. The ration of growth function is plotted in Fig.(\ref{Fig-D}).
In the next subsection we will introduce galaxy growth rate parameter as a new observational parameter.
%
%
%
\begin{figure}[t]
\centering
\includegraphics[width=10cm]{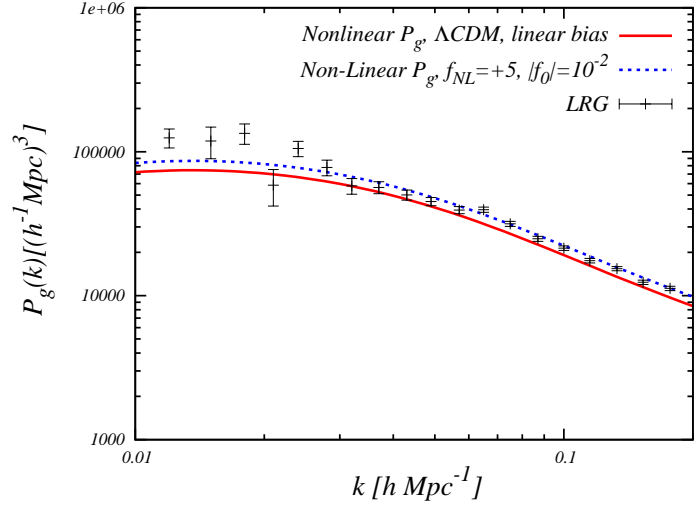}
\caption {The galaxy power spectrum is plotted versus redshift for $\Lambda$CDM case with linear bias (red-solid line) and for the modified gravity theory with the deviation parameter $|f_0|=10^{-2}$ and the local NG with $f_{NL}=+5$ (blue-dotted line) with non-linear corrections. The data points are the galaxy power spectrum from the Luminous Red Galaxy (LRG) data from SDSS survey \cite{Tegmark:2006az}.} \label{Fig-Pg}
\end{figure}


%
%
\begin{figure}[t]
\centering
\includegraphics[width=10cm]{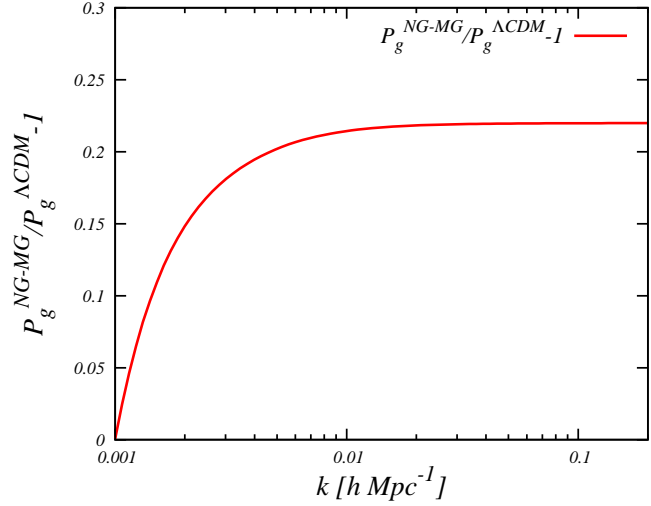}
\caption {The ratio of the galaxy power spectrum for the modified gravity theory with the deviation parameter $|f_0|=10^{-2}$ and the local NG with $f_{NL}=+5$ and $\Lambda$CDM case with linear bias. } \label{Fig-Pgrel}
\end{figure}


\subsection{Galaxy Growth Rate}

In  previous subsection we discussed the effect of MG and NG on RSD and growth rate of dark matter. Observationally we measure the RSD through the power spectrum or galaxy correlation function, where by knowing the bias parameter we will obtain the growth rate. Now we define a new parameter known as galaxy growth rate as:\
\be
f_{g}=\frac{d\ln\delta_g}{d\ln a}
\ee
where $\delta_g$ is the galaxy number density contrast. The $f_g$ parameter could be a direct observable if we have enough statistics of galaxies in redshift bins very close to each other to obtain differential quantity $d\delta_g/dz$. Now to go further we can relate the galaxy growth rate to the dark matter growth rate as follows:
\be
f_g(k,z)=\frac{d\ln(b\delta_m)}{d\ln a}=f_m(k,z)+\frac{d\ln b(k,z)}{d\ln a}=f_m-(1+z)\frac{b'(k,z)}{b(k,z)}
\ee
where $'$ denotes the derivative with respect to redshift.
In the case of linear bias, and the universality assumption, the bias term can be written as a function of height parameter $b=b(\nu)$. Consequently, the redshift dependence only appears  in $\sigma(M,z)$.
Accordingly, we can write the galaxy growth rate as:
\be
f_g(k,z)=f_m(k,z)-(1+z)\frac{d\ln b(\nu)}{d\ln \nu}\nu\delta_c\frac{d\sigma_M(z)}{dz}\sigma_M^{-2}(z)
\ee
Now using the fact that $\sigma_M(z)=\sigma_M(z=0){D(z)}/{D(z=0)}$ and $f_m=-(1+z)D'(z)/D(z)$ we will find:
\be{\label{Eq-fg1}}
f_g(k,z)=\left[1-\frac{d\ln b(\nu)}{d\ln(\nu)}\right]f_m(k,z)
\ee
Eq.(\ref{Eq-fg1}) shows that there is a linear bias between the growth rate of  galaxies and the growth rate of matter where we have defined  the {\it{linear growth rate bias}} as:
\be
b^{(L)}_{(f)}(z)\equiv\left[1-\frac{d\ln b(\nu)}{d\ln(\nu)}\right]
\ee
in which the superscript $(L)$ indicates the linearity of the growth rate bias. For the Press-Schechter and the Sheth-Tormen  mass function the growth rate bias is obtained as
\be
b^{L}_{PS(f)}(z)=\frac{\delta_c-1-\nu^2}{\delta_c-1+\nu^2}\
\ee
\be\label{Eq-bfST}
b^{L}_{ST(f)}(z)=1-\frac{2\alpha\nu^2}{\delta_c+\alpha\nu^2-1+\frac{2p}{1+(\alpha\nu^2)^p}}\left[1-2p^2 \frac{(\alpha\nu^2)^{p-1}}{\left(1+(\alpha\nu^2)^p\right)^2}\right]
\ee

In Eq.(\ref{Eq-bfST}), if  we set $p=0$ and  $\alpha=1$, not surprisingly, the galaxy growth rate bias of Sheth-Tormen will be the same as Press-Schechtre bias.
Now we turn our attention to the NG galaxy growth rate bias. In this case the bias will be a function of height and the cosmological evolution of perturbation via parameter ${\cal{M}}$ as:
\be
\tilde{b}^{(L)}_f(k,z)=\tilde{b}^{(L)}(\nu, {\cal{M}}(k,z))
\ee
If  the functionality of  $\nu$ and ${\cal{M}}$  are separable $\tilde{b}^{(L)}_f(k,z)=\bar{b}_{{\cal{M}}}[{\cal{M}}(k,z)]\times\bar{b}_{\nu}[\nu(z)]$ , the growth rate bias will be:
\be
\tilde{b}^{(L)}_f(k,z)=\left[b^{(L)}_f+\frac{1}{f_m}\frac{d\ln \bar{b}_{{\cal{M}}}[{\cal{M}}(z,k)]}{d\ln a}\right],
\ee
where a new term added to growth rate bias term which has a scale dependence.
In the case of the MG growth rate will be modified by introducing a ${\cal{M}}_{MG}(z,k)$, instead of the standard term.


\section{Conclusion and Discussion}

 In this work we study the simultaneous effect of the Hu-Sawicki $f(R)$ modified gravity theories and the local primordial NG on Large scale structure observables. We use the specific model of $f(R)$ just as a toy model to show the effect the potentially important effect of MG in interpretation of the LSS observations. The modification of the gravity, introduces a scale dependence in the growth of the structures. On the other hand the primordial NG make the bias parameter a scale dependent quantity. In order to study the effect of NG/MG  we assert  that  galaxy power spectrum and redshift space distortion are  promising observables. As both of them are affected by the bias parameter and  the growth rate of the structures. In the case of the modified gravity the bias parameter and growth rate function become scale dependent quantities. In this work we assume that the scale dependence in the linear bias parameter due to modification of the gravity is small and the new scale dependence comes from the contribution of modified growth rate in non-Gaussian bias.
We show that the  redshift space distortion parameter, in higher redshifts and larger scales can be used to distinguish between the positive and negative $f_{NL}$ (plotted in Fig.(\ref{Fig-beta})) where the background is $\Lambda$CDM.  However, in the case of $f(R)$ gravity with deviation of $|f_0|=10^{-2}$  and the existence of primordial NG, there is a degenerate effect in higher redshifts for RSD. On the other hand the galaxy power spectrum is affected by redshift space distortion and bias simultaneously, where the scale dependence shows up again. Generally introducing a primordial NG or a modification of gravity enhance the galaxy power spectrum(\ref{Fig-Pg}). This enhancement has a slight scale dependence shown in Fig.(\ref{Fig-Pgrel}). This scale dependence is mainly sourced by the modification of the gravity than the bias parameter, consequently we show that this scale dependence shows up in smaller scales. Finally, in order to have a new cross-check for our observations and breaking the degeneracies,  we introduce the galaxy growth rate. The galaxy growth rate which is a measure of the growth of the galaxies with respect to scale factor in logarithmic scale  lead us to the growth rate bias parameter. The growth rate bias parameter relates the galaxy growth rate to dark matter growth rate. In the $\Lambda$CDM case this galaxy growth rate bias parameter is scale independent, while by introducing the NG/MG, the scale dependence emerged.

It is worth to mention that future observation of large scale structure, mainly the galaxy counting with better precision and larger statistics will help us to constrain the cosmological models with deviation from six parameter standard model. The galaxy power spectrum, redshift space distortion and bias parameter is potentially promising observables to detect any scale dependence in them. Although there are many complications in the detection of departure from $\Lambda$CDM, where we list them as below with corresponding discussions: a) The detection of small NG in LSS observables (i.e. bias parameter)
is a difficult task due to the statistics and the noise. However there are optimistic forecasts for future. On the hand there is a room for {\it{scale dependence bias}}, where we can get larger NG in sub-CMB scales.
b)Constraining the cosmological models with galaxy power spectrum(correlation function) in small scales is always a venturesome task because of non-linear effects. The deviation from linear power spectrum, which is most affected with primordial NG and background evolution, could be a result of the non-linear growth of the structures. The future very large scale surveys can probe the linear power spectrum.
c) The interpretation of the bias parameter is a complicated task. As the bias parameter depend on the sample of luminous matter that we choose (i.e. the color, mass, redshift ...  of the galaxy sample). Consequently the existence of the degeneracy between the astrophysical effects and the cosmological ones are there.
d)The scale dependence of linear bias because of the scale dependent growth function will be a source of uncertainty as well.
As a future prospect of this work, it is possible to extend the study of the  simultaneous effect of the general NG shape with the general modified growth function. Also  future LSS observations will provide statistically meaningful data to constrain the $f_{NL}$ and deviation from $\Lambda$CDM parameter simultaneously.
\label{Sec-Conc}

\acknowledgments
We would like to thank Sarah Shandera, Hassan Firouzjahi and Sohrab Rahvar for their insightful comments and discussions. Also we want to thank anonymous referee for his/her useful comments which help us to improve our representation and results. NM  thanks the School of Astronomy of Institute for Research in Fundamental Science (IPM) for their kind hospitality during the preparation of this work.

\end{document}